\documentclass[aps,pra,groupaddress,showpacs,twocolumn]{revtex4}
\usepackage{units}
\usepackage{amsmath}
\usepackage{amssymb}
\usepackage{graphicx}
\usepackage{bm}
\usepackage{color}

\newcommand{\be}{\begin{equation}}
\newcommand{\ee}{\end{equation}}
\newcommand{\bea}{\begin{eqnarray}}
\newcommand{\eea}{\end{eqnarray}}
\newcommand{\Sc}{S_c}

\newcommand{\pt}{\mathcal{PT}}

\begin{document}
\title{Conservation relations and anisotropic transmission resonances in one-dimensional $\pt$-symmetric photonic heterostructures}

\author{Li Ge}
\affiliation{Department of Electrical Engineering, Princeton University, Princeton, New Jersey 08544}
\author{Y.~D.~Chong}
\affiliation{Department of Applied Physics, Yale University, New
  Haven, Connecticut 06520}
\author{A.~D.~Stone}
\affiliation{Department of Applied Physics, Yale University, New
  Haven, Connecticut 06520}


\date{December 11th, 2011}

\begin{abstract}
We analyze the optical properties of one-dimensional (1D) $\pt$-symmetric structures of arbitrary complexity.
These structures violate normal unitarity (photon flux conservation) but are shown to satisfy generalized unitarity
relations, which relate the elements of the scattering matrix and lead to a conservation relation in terms of the transmittance
and (left and right) reflectances. One implication of this relation is that there exist anisotropic transmission resonances
in $\pt$-symmetric systems, frequencies at which there is unit transmission and zero reflection, but only for waves incident from
a single side. The spatial profile of these transmission resonances is symmetric, and they can occur even at $\pt$-symmetry breaking points. The general conservation relations can be utilized as an experimental signature of the presence of $\pt$-symmetry and of $\pt$-symmetry breaking transitions. The uniqueness of $\pt$-symmetry breaking transitions of the scattering matrix is briefly discussed by comparing to the corresponding non-Hermitian Hamiltonians.
\end{abstract}

\maketitle

\section{Introduction}

Motivated by fundamental studies of $\pt$-symmetric quantum Hamiltonians \cite{Bender1,Bender2,Bender3}, $\pt$-symmetric photonic structures
have attracted considerable interest in the past few years.  These are structures with balanced gain and loss; in the case of a
one-dimensional (1D) structure, this means that there is a symmetry point (chosen to be the origin, $x=0$) around which the linear index of refraction satisfies $n^* (-x) = n(x)$.  Such structures were first studied in Refs.~\cite{El-Ganainy_optlett07,Makris_prl08} and were shown to exhibit a variety of exotic photon transport phenomena, such as double refraction \cite{Makris_prl08}, power oscillations \cite{Makris_prl08,ruter_natphy10,Zheng_pra10}, and non-monotonic behavior of the transmission loss with increased dissipation \cite{Guo_prl2009}. The initial studies focused on parallel waveguide structures with alternating loss and gain, in which the transverse variation of the electrical field, in the paraxial approximation to the wave equation, maps precisely onto a 1D or discrete Schrodinger equation, similar to the earlier quantum studies \cite{El-Ganainy_optlett07,Makris_prl08,ruter_natphy10,Zheng_pra10,Guo_prl2009,Musslimani_prl08}. The parallel waveguide realization of $\pt$ symmetric photonic structures has recently found a promising application to compact optical isolators and circulators \cite{Ramezani}.

Recently, several authors have studied $\pt$-symmetric cavities
and heterostructures \cite{mostafazadeh,Schomerus,Longhi,CPALaser}, as
well as general $\pt$ scattering systems \cite{CPALaser}, using the full scalar
wave equation, in the case that it obeys at least one $\pt$ symmetry operation.
The current authors in particular emphasized the existence in such
systems of $\pt$-symmetric and $\pt$-broken phases of the
electromagnetic scattering matrix (S-matrix). For the 1D case, the
eigenvalues of the S-matrix are unimodular in the $\pt$-symmetric
phase, as they are in unitary systems, but photon flux is
$\textit{not}$ conserved for most scattering processes, whereas in the
$\pt$ broken phase the S-matrix eigenvalues have reciprocal
magnitudes, one greater than unity (corresponding to amplification),
and the other less than unity (corresponding to attenuation).  We and
others \cite{Longhi,Schomerus,CPALaser,mostafazadeh,schomerus2} pointed out the
existence of novel singular points in the broken symmetry phase, which
we refer to as CPA-laser points.  At these points one of the S-matrix
eigenvalues goes to infinity (the usual lasing threshold condition),
while the other goes to zero.  The latter phenomenon corresponds to
coherent perfect absorption (CPA)\cite{CPA,CPA_science}, in which a
specific mode of the electromagnetic field, the time-reversal of the
lasing mode, is completely absorbed.  For $\pt$-symmetric
structures, these two phenomena must coincide \cite{Longhi,CPALaser};
i.e.~at the laser threshold, in addition to a radiating mode of
self-oscillation, there always exists an incident field pattern which,
instead of being amplified, is completely attenuated.

The rich behavior of 1D $\pt$-symmetric photonic structures
violates the standard intuition that optical structures can be
characterized by their single-pass gain or loss, which is always zero
in these systems. The coincidence of both lasing and perfect
absorption, and more generally the reciprocal amplification and
attenuation displayed by the S-matrix eigenvalues, is a strict
consequence of the symmetry property of the S-matrix for such
structures.  In Ref.~\cite{CPALaser} this was expressed in arbitrary
dimensions by the relation
\begin{equation}
  (\pt) \, S(\omega^*) \, (\pt) = S^{-1}(\omega), \quad
  \label{S matrix PT}
\end{equation}
where $\cal P$ is the parity operator (or indeed any discrete symmetry
operator with ${\cal P}^2 =1$), and $\cal T$ is the time-reversal
operator (in the representation we will employ, this can be taken as
the complex conjugation operator).  By comparison, a $\cal
T$-symmetric unitary S-matrix would obey ${\cal T} S(\omega^*) {\cal
  T} = S^{-1}(\omega)$.

The set of S-matrices obeying Eq.~(\ref{S matrix PT}) can be shown to be isomorphic
to a pseudo-unitary group, which in the 1D case is just $U(1,1)$ \cite{Mostafazadeh2}.
In physical dimensions
higher than one, there can be more than two input and output channels,
and it is possible for the S-matrix to be in a mixed ``phase'' with
one subset of the eigenvalues forming ``$\mathcal{PT}$-broken''
amplifying/attenuating pairs and the remaining eigenvalues being
``$\mathcal{PT}$-symmetric'' and flux conserving.  For 1D structures,
however, there are only two eigenvalues, and they must either be both
unimodular, or a non-unimodular inverse conjugate pair---except at the
$\pt$-transition point, an exceptional point at which the S-matrix has
only one eigenvector and eigenvalue \cite{CPALaser}.

Several specific cases of 1D $\pt$-symmetric structures have been
studied \cite{mostafazadeh,Longhi,CPALaser,invisibility} and in
addition to the interesting CPA-laser behavior, other intriguing
properties have been found, such as unidirectional invisibility
\cite{invisibility}.  It is thus worthwhile to see what specific
properties $\pt$-symmetry imposes on transmission and reflection in
arbitrary $\pt$ structures, in both the symmetric and broken-symmetry
phases. That is the goal of the current work.  In Section
\ref{Generalized Unitarity Relations}, we show that 1D $\pt$
structures obey certain strong conservation relations, which could be
employed experimentally to determine if a given structure has realized
$\pt$ symmetry.  In Section \ref{Anisotropic Flux-Conserving
  Transmission Resonances}, we examine a consequence of these
conservation relations: the existence of transmission resonances in
which the reflectance vanishes only for waves incident from one side
of the structure, which we refer to as anisotropic transmission
resonances (ATRs).  The uni-directional invisibility
phenomenon found by Lin {\it et al.}~\cite{invisibility} is a special case of
these ATRs.  In Section
\ref{Phase transition boundaries}, we derive a separate relation for
the boundary between the $\pt$-symmetric and $\pt$-broken phases of
the S-matrix, involving the reflectance and transmittance for
one-sided scattering processes.  In Section \ref{Uniqueness of PT
  transition in scattering}, we show that our conventional definition
of the S-matrix and its eigenvalues is physically meaningful, and in
particular that its phase boundary can be related to $\pt$-breaking
transitions in the spectrum of some $\pt$-symmetric Hamiltonian.

\section{Generalized Unitarity Relations}
\label{Generalized Unitarity Relations}

\begin{figure}
  \centering
  \includegraphics[width=0.8\linewidth]{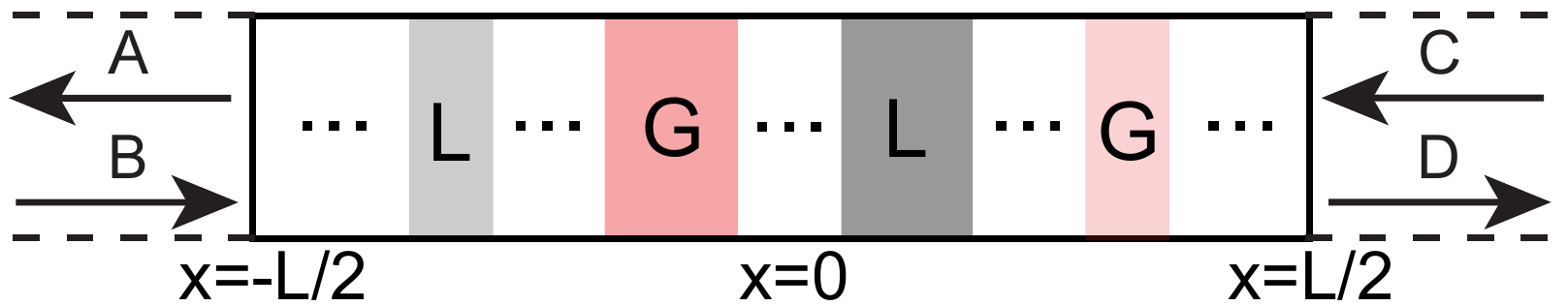}
  \caption{Schematic of a 1D $\pt$-symmetric photonic
    heterostructure, consisting of an arbitrary number of layers that
    are $\pt$-symmetric about $x=0$, i.e.~$n(x)=n(-x)^*$. G (red)
    and L (gray) indicate gain and loss regions, and different color
    tones indicate different amplification/absorption
    strengths.} \label{fig:schematic}
\end{figure}

We begin, following Longhi \cite{Longhi}, with the 1D transfer matrix
$M$, defined by (see Fig.~\ref{fig:schematic}):
\begin{equation}
\left( \begin{matrix} A\\ B\end{matrix}\right) = M
  \left(\begin{matrix} C\\ D\end{matrix}\right).
\end{equation}
For a $\pt$-symmetric heterostructure, the components of $M$ obey
the following properties \cite{Longhi}:
\be M_{22}(\omega)=M_{11}^*(\omega^*),\,
M_{12(21)}(\omega)=-M_{12(21)}^*(\omega^*),\label{eq:M0} \ee
where $\omega$ is the frequency of the incident/scattered beams. For
real $\omega$, these relations imply $M_{22}=M_{11}^*$ and
$\text{Re}[M_{12}]=\text{Re}[M_{21}]=0$, which enables us to
parameterize $M$ as
\be M = \left(\begin{array}{c c} a^* & ib \\ -ic & a \end{array}\right).
\label{eq:M}
\ee
It is determined by three independent {\it real} quantities, i.e. $b$
and the phase and amplitude of $a$.  The parameter $c$ is related to
$|a|,b$ by
\begin{equation}
  b c = |a^2|-1,
  \label{eq:c_constraint}
\end{equation}
which arises from the quite general condition $\textrm{det}(M)=1$
\cite{Yeh}. The parametrization using $a,b$ is valid except when
$M_{12}=0$; in that case, $|a|=1$ and $c$ replaces $b$ as the third
independent parameter.

In the following discussion we assume non-vanishing $M_{11}$ and
$M_{22}$, which holds everywhere except at CPA-laser points
\cite{bibnote0}. The S-matrix
is defined by
\begin{equation}
\left( \begin{matrix} A\\ D\end{matrix}\right) = S \left( \begin{matrix} B\\ C\end{matrix}\right) \equiv \left( \begin{matrix} r_L& t\\ t& r_R\end{matrix}\right) \left( \begin{matrix} B\\
C\end{matrix}\right), \label{eq:Sdef}
\end{equation}
where $r_{L/R}$ are the reflection coefficients for light incident
from the left and right respectively, while $t$ is the transmission
coefficient, which is independent of the direction of incidence.  The
parametrization (\ref{eq:M}) gives
\be
S = \frac{1}{a}\left(\begin{array}{c c}
        ib & 1 \\
        1 & ic
       \end{array}\right). \label{eq:S}
\ee
Thus the reflection coefficients are $r_L=ib/a$ and $r_R = ic/a$,
which are unequal in magnitude but can differ in phase by only 0 or
$\pi$; and the transmission coefficient is $t = 1/a$.  Note that $S$
satisfies the symmetry relation (\ref{S matrix PT}), with $\mathcal{P}
= \left(\begin{smallmatrix}0 & 1\\1 & 0 \end{smallmatrix}\right)$ and
$\mathcal{T}$ the complex conjugation operator.  Using (\ref{eq:c_constraint}), we obtain the following exact ``generalized unitarity
relation'':
\be r_Lr_R = t^2 \left(1-\frac{1}{T}\right). \label{eq:constraint0}
\ee
This leads to the conservation relation
\be |T-1| = \sqrt{R_LR_R}, \label{eq:constraint} \ee
where $R_{L/R} \equiv |r_{L/R}|^2$ are the two reflectances and $T
\equiv |t|^2$ is the transmittance.  In addition,
Eqs.~(\ref{eq:S},\ref{eq:constraint0}) lead to phase relationships
among the reflection and transmission coefficients:
\begin{align}
  \begin{aligned}
 &\phi_R = \phi_L, & \text{if}\; T<1; \\
 &\phi_R = \phi_L +  \pi,& \text{if}\; T>1;\\
 &\phi_{L,R} = \phi_t \pm \pi/2,
  \end{aligned}\label{eq:constraint_phase}
\end{align}
where $\phi_{L,R},\phi_t$ are the phases of the reflection and transmission coefficients.

\begin{figure}[htbp]
  \centering
  \includegraphics[width=7cm]{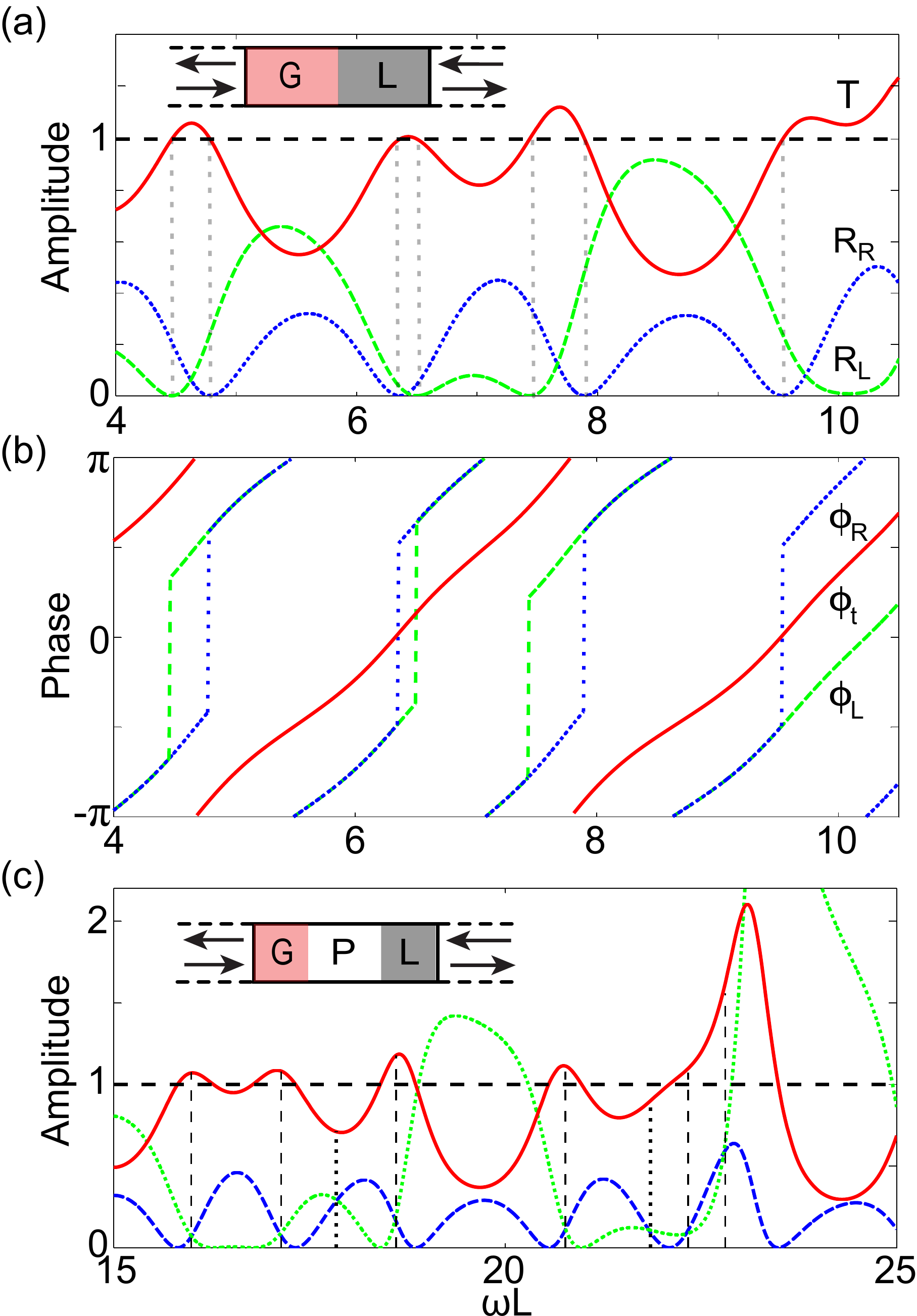}
  \caption{(a) Reflectance and transmittance of a 1D $\pt$-symmetric structure of index $n=2\pm0.2i$ and length $L$. $R_L,R_R,T$ are indicated by the red, green, blue curves. Zeros of the reflectances and the corresponding anisotropic transmission resonances ($T=1$) are marked by dotted grey lines. The quantity $R_LR_R + 2T-T^2=1$ is plotted as the dashed black line to demonstrate the conservation relation (\ref{eq:constraint}). (b) Phases of $r_L$, $r_R$, and $t$ in (a), demonstrating the reflection phase jumps at each ATR.  (c) Same quantities are plotted as in (a), but the structure now has a passive region of length 2L/5 in the center. For this structure we see that it is possible to have ``accidental" flux-conserving points at which
$R_L = R_R (\equiv R\neq0)$ and hence $T+R=1$.  Similarly, there are accidental pseudo-unitary points for which $T-R=1$. They are indicated by vertical dotted and dashed lines, respectively.} \label{fig:RT}
\end{figure}

Eqs.~(\ref{eq:constraint0},\ref{eq:constraint}) are the central
results of this work. They are valid for all 1D photonic
heterostructures with $\pt$-symmetry; two examples are shown in
Fig.~\ref{fig:RT}.

For $T<1$, Eq.~(\ref{eq:constraint}) becomes $T+\sqrt{R_LR_R}=1$.
This is an intriguing generalization of the more familiar conservation
relation $R + T = 1$, which applies to unitary
($\mathcal{T}$-symmetric) S-matrices for which the left and right
reflectances are necessarily equal.  In the $\mathcal{PT}$-symmetric
case, the geometric mean of the two reflectances, $\sqrt{R_LR_R}$,
replaces the single reflectance $R$.  Therefore, when $T<1$, the
scattering of a single incident wave from one side of the structure is
sub-unitary (some flux is lost) and the scattering from the other side
is super-unitary (some flux is gained).  As an exception, there can be
an accidental degeneracy at which $R_L = R_R$, in which case the
scattering from both sides conserves flux.  Such special cases do occur as
a continuous parameter such as frequency is varied
for non-trival $\pt$ systems ($\text{Im}[n(x)] \neq 0$), as shown in
Fig.~\ref{fig:RT}(c).

For $T>1$, all single-sided scattering processes are super-unitary,
and the conservation relation (\ref{eq:constraint}) can be re-written
as $T- \sqrt{R_LR_R}=1$.  Accidental reflectance degeneracies ($R_L =
R_R$) are also possible in this regime, giving the usual pseudo-unitary
conservation relation $T-R=1$, as shown in
Fig.~\ref{fig:RT}(c). All of these quantities actually diverge when approaching the CPA-laser points, but they still satisfying the conservation relation (\ref{eq:constraint}).

Finally, we see that for $T=1$, one of the reflectances must vanish
(the other typically does not).  Hence, the scattering for that
direction of incidence is flux conserving, similar to resonant
transmission in unitary structures.  This phenomenon is analyzed in
greater detail in Section \ref{Anisotropic Flux-Conserving
  Transmission Resonances}.

Interestingly, the S-matrix describing
three-wave mixing in the undepleted pump approximation corresponds to the the special case where $R_L=R_R$
\cite{Longhi2,Stone_unpub}. The case $T+R = 1$ describes frequency conversion by absorption or emission of a pump photon, and $T-R = 1$ describes parametric amplification of both signal and idler by down conversion of pump photons.  The relevance of a special case of $\pt$ symmetry to optical parametric amplification and conversion has only very recently been appreciated.

An experimental concern in all $\pt$ systems is how to confirm that
one has truly realized a structure with $\pt$-symmetry, i.e.~that the
gain and loss are balanced and the real index is symmetric.
Eqs.~(\ref{eq:constraint0}-\ref{eq:constraint_phase}) are strong
constraints on the allowed scattering processes with a single incident
beam for $\pt$ systems, and can be used to test how close one is to
the ideal symmetric structure.

\section{Anisotropic Flux-Conserving Transmission Resonances}
\label{Anisotropic Flux-Conserving Transmission Resonances}

\begin{figure}
  \centering
  \includegraphics[width=\linewidth]{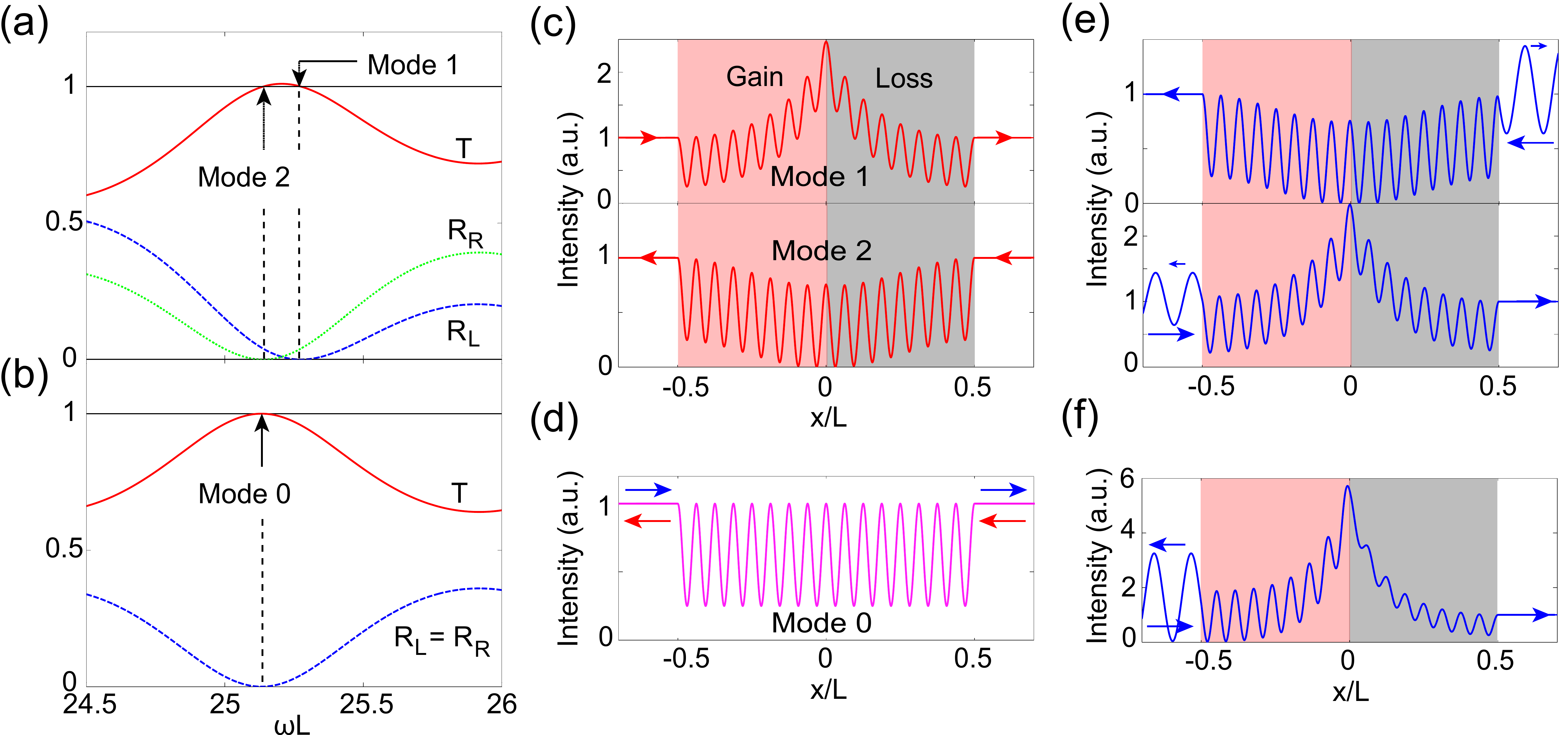}
  \caption{Asymmetric transmission resonances in a multi-layer heterostructure. (a) Transmittance and reflectances as a function of frequency
 for $\text{Im}[n ]\neq 0 $ and, (b), for $\text{Im}[n]=0$. Mode 0 is a bi-directional transmission resonance in the latter case. The structure has a constant $\text{Re}[n]=2$ and consists of 50 layers with $\text{Im}[n]$ increasing (decreasing) stepwise from $0.004$ ($-0.004$) to $0.1$ ($-0.1$) towards the center in the loss (gain) half. The frequencies are $\omega L=25.275,\,25.139$ in Mode 1 (wave incident from gain side) and 2
(wave incident from the loss side). Spatial profiles shown are: (c) the ATRs, (d) unitary bi-directional transmission resonance, (e) ``wrong'' side scattering at the ATR frequencies, and (f) ``wrong" side scattering at $R_R=0$ in the same structure but with larger $\text{Max[Im}[n]]=0.17$, showing a stronger asymmetry.} \label{fig:tunnelingState}
\end{figure}

As we have noted, Eq.~(\ref{eq:constraint}) implies an interesting
phenomenon: there exists a flux-conserving scattering process for
incident waves on a single side if and only if $T=1$, and one of $R_L$
or $R_R$ vanishes.  We refer to such a process as an anisotropic
transmission resonance (ATR).  ATRs are different from the accidental
flux-conserving processes that can occur for $T<1$; those, as we have
seen, are accessible from either direction of incidence ($R_L=R_R$).
ATRs are a generalization of the flux-conserving transmission
resonances of unitary systems, which are independent of the incidence
direction.  In Fig.~\ref{fig:tunnelingState}, we show how two ATRs
evolve out of a single transmission resonance of the unitary system as
balanced gain and loss is added.  Within the same structure, ATRs can
occur for both left and right incidence, as the frequency is varied, but generally at different
frequencies (to occur at the same frequency, a ``doubly accidental''
degeneracy $R_L=R_R = 0$ would have to occur, requiring a second tuning parameter).

A surprising property of ATRs is that their intensity profile is
spatially symmetric. This can be shown from the following analysis.
If $E(x)$ is the spatial profile of a left-/right-going transmission
resonance, then by a $\pt$ operation $E^*(-x)$ is also a
left-/right-going transmission resonance of the same structure. Since
these two states happen at the same frequency, they must be identical
(up to a phase $\phi$) by the requirement of uniqueness:
\be E^*(-x) = e^{i\phi}E(x). \label{eq:symm} \ee
Hence, the intensity satisfies $I(x)\equiv|E(x)|^2=I(-x)$. This result
is consistent with the intuitive expectation that in order to conserve
flux the photons must on average spend equal amount of time in the
loss and gain regions of the structure.  Except at the ATRs,
intensities do exhibit asymmetry for single-sided incidence, and in
particular this is the case for a wave incident from the side with
non-vanishing reflectance; see Fig.~\ref{fig:tunnelingState}(e),(f).

Fig.~\ref{fig:tunnelingState} shows two ATRs of a multi-layer
structure, one for each incidence direction, occurring at different
frequencies.  The frequencies are very similar because $\text{Im}[n]$
is not very large and both ATRs arise from a bi-directional
transmission resonance of the unitary ($\text{Im}[n]=0$)
heterostructure. As we add gain and loss to the unitary
heterostructure, while preserving the $\pt$-symmetry, the transmission
resonances separate and their spatial profiles become more distinct.
Fig.~\ref{fig:tunnelingState}(e) shows the asymmetric intensity
profiles for waves incident at the ATR frequency, but from the
``wrong'' side (the side with non-vanishing reflectance).  The
asymmetry increases as the two ATRs move further apart with increasing
gain/loss, as shown in Fig.~\ref{fig:tunnelingState}(f).

Let us refer to the left and right halves of a $\pt$-symmetric
heterostructure as $U,V$.  We can write the reflection and
transmission coefficients coefficients for the whole structure ($r_L$,
$r_R$, and $t$) in terms of the reflection and transmission
coefficients for the $U$ and $V$ segments:
\bea
r_L &=& \frac{r_{L,U}-e^{i(\alpha_U+\alpha_V)} r^*_{L,U}}{1 - e^{i(\alpha_U+\alpha_V)}r_{L,U}^*r_{R,V}^*}, \label{eq:rL}\\
r_R &=& \frac{r_{R,V}-e^{i(\alpha_U+\alpha_V)} r_{R,V}^*}{1 - e^{i(\alpha_U+\alpha_V)}r_{L,U}^*r_{R,V}^*}, \label{eq:rR}\\
t &=& \frac{e^{i\alpha_U}(1-r_{L,U}^*r_{R,V})}{1 - e^{i(\alpha_U+\alpha_V)}r_{L,U}^*r_{R,V}^*}. \label{eq:t}
\eea
Here, $\alpha_{U/V} \equiv 2\text{Arg}[t_{U/V}]$.  Note that if either
$r_{L,U} = 0 $ or $r_{R,V} = 0$ at some $\omega$, corresponding to a
transmission resonance of $U/V$ in right/left direction, the
transmittance for the full structure will also be unity.

\begin{figure}
  \centering
  \includegraphics[width=0.8\linewidth]{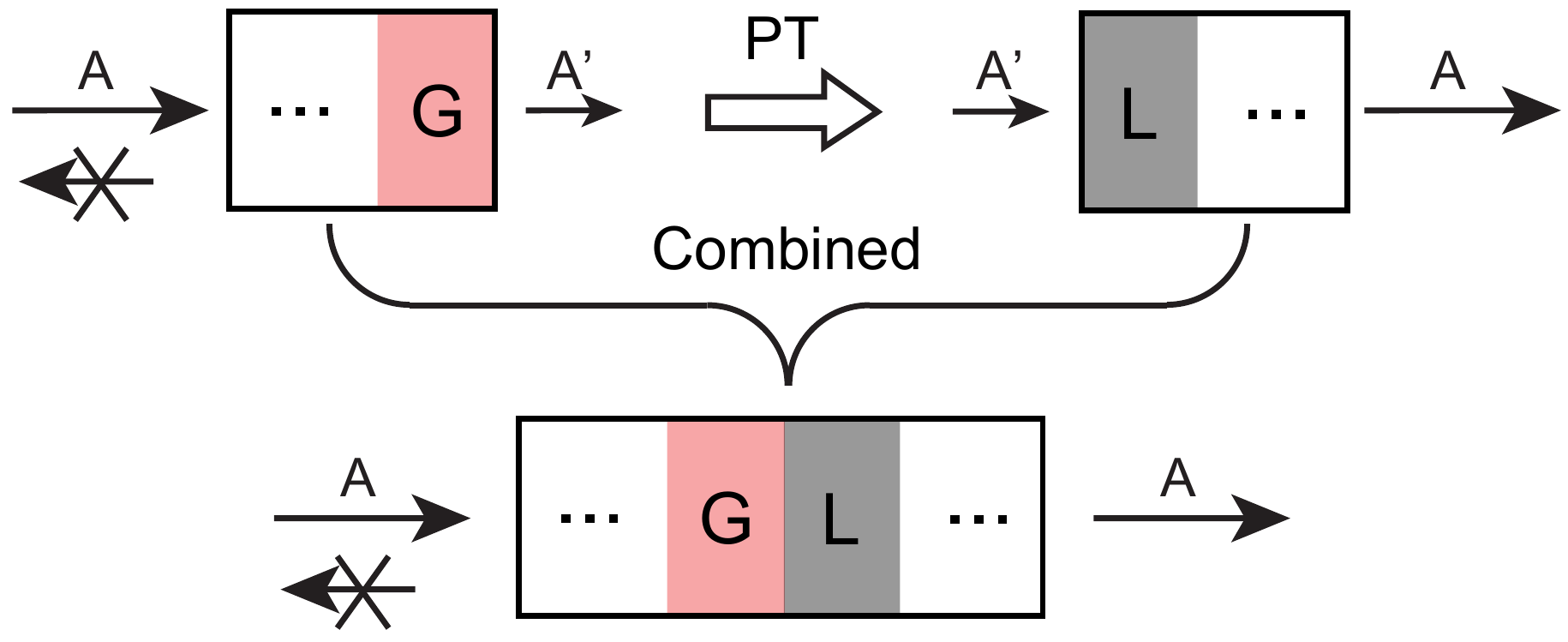}
  \caption{Graphical explanation of a trivial ATR arising from the transmission resonance of the left half in the right direction.
  $\pt$ symmetry requires that it also be a resonance of the right half for left incidence (see argument in text).
  Depending on the particular $\pt$ structure there may or may not be trivial ATRs; for example the simple heterostructure
  of Fig.~\ref{fig:RT}(a) has no such ATRs.  The ATRs of primary interest are those that arise from multiple scattering between the
  left and right halves of the structure.} \label{fig:trivialATRs}
\end{figure}

Thus one type of ATRs can arise from resonances of either {\it half} of the $\pt$ system. This follows from
$\pt$ symmetry.  First, using the time-reversal operation, a transmission resonance of $S(nk)$ from the left
must be a transmission resonance of $S(n^*k)$ from the right (interchange gain and loss regions and interchange
incoming and outgoing amplitudes)\cite{CPA_science}.  Second, the S-matrix of the right hand side of a $\pt$ structure
is ${\cal P} S(n^*k)$, so the right half of the $\pt$ structure must have a resonance for waves incident
from the left side as well, if its left side does.  Therefore the composite structure will have an ATR if either
half does ($r_{L,U} = 0 $ or $r_{R,V} = 0$).  This argument is illustrated graphically in Fig.~\ref{fig:trivialATRs}; we refer to these as trivial ATRs.

ATRs also occur when $\text{Arg}[r_{L,U}]$ or $\text{Arg}[r_{R,V}]$ equals $(\alpha_U+\alpha_V)/2$
and involve multiple scattering between the sub-units. It is straightforward to check that at such points $T=1$ and $R_L(R_R)=0$. It can be shown that a single layer of gain or loss in a lossless environment (e.g. in air) does not have transmission resonances in general, and we show in the appendix that all the ATRs in Fig.~\ref{fig:RT} are of this type and are thus ``non-trivial".

As already noted, for an ATR to be bi-directional, a doubly accidental degeneracy is
needed either in the amplitude of $r_{L,U}$ and $r_{R,V}$
($r_{L,U}=r_{R,V}=0$) or their phase
($\text{Arg}[r_{L,U}]=\text{Arg}[r_{R,V}]=(\alpha_U+\alpha_V)/2$).
This is highly unlikely, unless one can tune an additional continuous parameter other
than the frequency, so in the generic case all transmission resonances of $\pt$
structures are uni-directional.

In a recent work, Lin {\it et al.}~\cite{invisibility} have studied a 1D $\pt$-symmetric Bragg structure of alternating dielectric layers with appropriate gain and loss, and discovered an ATR with an additional property which they refer to as ``uni-directional invisibility". Not only do they find $T=1, R_L = 0, R_R \neq 0$ (or vice versa), as dictated by Eq.~(\ref{eq:constraint}); they also find that at this ATR the transmission phase $\phi_t = 0$, corresponding to zero phase delay of the signal compared to free propagation.  For this reason there would be no signature of the presence of the structure in either amplitude or phase of the received wave, if the wave is sent from the correct side (there would be a signature of course in the reflected wave if sent from the wrong side).  This condition, that $\phi_t = 0$ at the ATR, is not required by our generalized unitarity relations and is specific to their structure \cite{bibnote:accidental_invisibility}.

The existence of non-trivial ATRs is independent of whether the S-matrix is in the $\pt$-symmetric or $\pt$-broken phase \cite{CPALaser}; They can even occur at the symmetry-breaking exceptional point (see the following section). However, we do find for the simple gain-loss heterostructure of Fig.~\ref{fig:RT}(a),(b) that the ATRs disappear soon after the lasing threshold is passed in the broken symmetry phase, since in the large $\text{Im}[n]kL$ limit $R_L R_R$ approaches unity asymptotically. This is not the case for more complicated $\pt$ structures such as that of Fig.~\ref{fig:RT}(c).  The different behaviors of the two cases are illustrated in Fig.~\ref{fig:r_gain} of the Appendix and its origin is discussed.

\section{Phase transition boundaries}
\label{Phase transition boundaries}

\label{sec:phaseBD}

A 1D $\pt$-heterostructure can undergo a spontaneous symmetry-breaking
transition in the eigenvalues and eigenvectors of its S-matrix, as
either $\omega$ is increased at fixed gain/loss or as gain/loss is
increased at fixed $\omega$ \cite{CPALaser}.  In the symmetric phase,
the $\pt$ operation maps each scattering eigenstate back to itself,
whereas in the broken symmetry phase each scattering eigenstate is
mapped to the other.  At the symmetry breaking exceptional point,
there is only one eigenvector and so both cases coincide.

Let $\lambda_{1,2}$ be the eigenvalues the S-matrix of a
$\pt$-symmetric heterostructure and $\nu_{1,2}$ be the ratios of the
two amplitudes of the corresponding eigenstates.  It follows from the
S-matrix parameterization (\ref{eq:S}) that
\bea
\lambda_1,\lambda_2 &=& \frac{i}{2a}\left[(b+c) \pm \sqrt{(b-c)^2-4}\right],\label{eq:eigCond} \\
\nu_1, \nu_2 &=& \frac{i}{2}\left[(c-b) \pm \sqrt{(b-c)^2-4}\right].\label{eq:vesCond}
\eea
These equations imply that $\lambda_1\lambda_2 = \nu_1\nu_2=-1$, and
the eigenvalues must have reciprocal moduli.  In the symmetric phase,
both eigenvalues are unimodular, whereas the broken symmetry phase
corresponds to the $|\lambda_1| > 1,| \lambda_2| < 1$ case.  The
exceptional point occurs when $b-c=\pm 2$, and there is a single
eigenvector with eigenvalue $\lambda = e^{\pm i\pi/2}$.  Both the
eigenvalues and amplitudes $\nu_{1,2}$ meet and bifurcate at the
exceptional point, similar to the $\pt$-breaking transitions
which occur in the eigenvalue spectra of $\pt$-symmetric
Hamiltonians \cite{Bender1,Bender2,Bender3,El-Ganainy_optlett07,Makris_prl08}.

\begin{figure}[h]
  \centering
  \includegraphics[width=7cm]{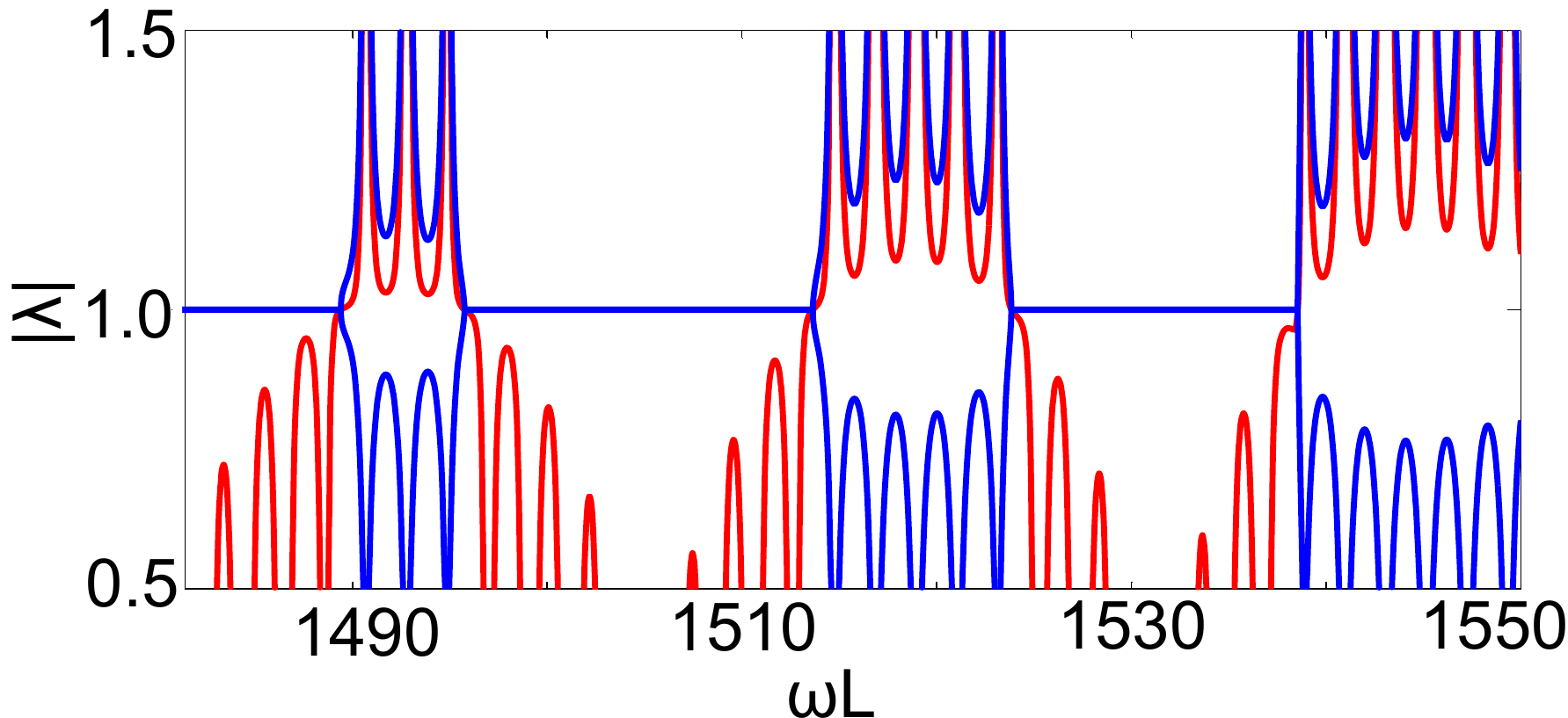}
  \caption{Test of the criterion (\ref{eq:borderCond}) for
    $\pt$-symmetry breaking points of a three-layer heterojunction
    structure. The blue lines represent the eigenvalues of the
    S-matrix, which exhibit five symmetry-breaking points as the
    frequency $\omega$ is tuned over the selected range.  The red line
    indicates the left hand side of Eq.~(\ref{eq:borderCond}). The
    heterostructure has a constant $\text{Re}[n]=3$, and the first and
    last layers are filled with gain and loss of Im$[n]=\pm0.005$. The
    width of the central passive region is $4\%$ of the total length
    $L$. } \label{fig:phasebd_3layers}
\end{figure}

Each eigenvector of the S-matrix corresponds to a particular choice of
two coherent beams, simultaneously directed at each side of the
heterostructure.  The S-matrix transition can in principle be observed
by tuning the complex input amplitudes, measuring the output
amplitudes, and hence finding the scattering eigenvalues.  One would
actually need to do such ``two-sided" interference experiments to detect the attenuating mode in
the broken symmetry phase, an interesting possibility which is
currently being explored \cite{Stone_unpub}.  However, such experiments with
two coherent input beams \cite{CPA_science} are often inconvenient and difficult to perform.
Therefore it would be preferable to have a criterion for the
transition based on separate single-beam measurements.

In Ref.~\cite{CPALaser} two such criteria were given for the phase boundaries in an arbitrary $\pt$-symmetric heterostructure;
however they both involve the relative phase of the reflection and transmission coefficients. One of these conditions is
$r_L-r_R=\pm 2it$.  Using the conservation relations (\ref{eq:constraint}) this can be shown to lead to the simpler condition \cite{bibnote:altExp}
\be
\frac{R_L+R_R}{2}-T=1,\label{eq:borderCond}
\ee
which involves only the transmittance and reflectances.  The left hand
side of Eq.~(\ref{eq:borderCond}) is greater than unity in the
broken-symmetry phase and less than unity in the $\pt$-symmetric
phase. This provides a simple experimental criterion for locating the
$\pt$-breaking transition point in 1D heterostructures.  This
criterion will be particularly useful if the quantity $(R_L + R_R)/2 -
T$ varies rapidly near the transition point.  This appears to be the
case for many heterostructures, as shown for example in
Fig.~\ref{fig:phasebd_3layers} for a three layer structure.

\begin{figure}[h]
  \centering
  \includegraphics[width=7cm]{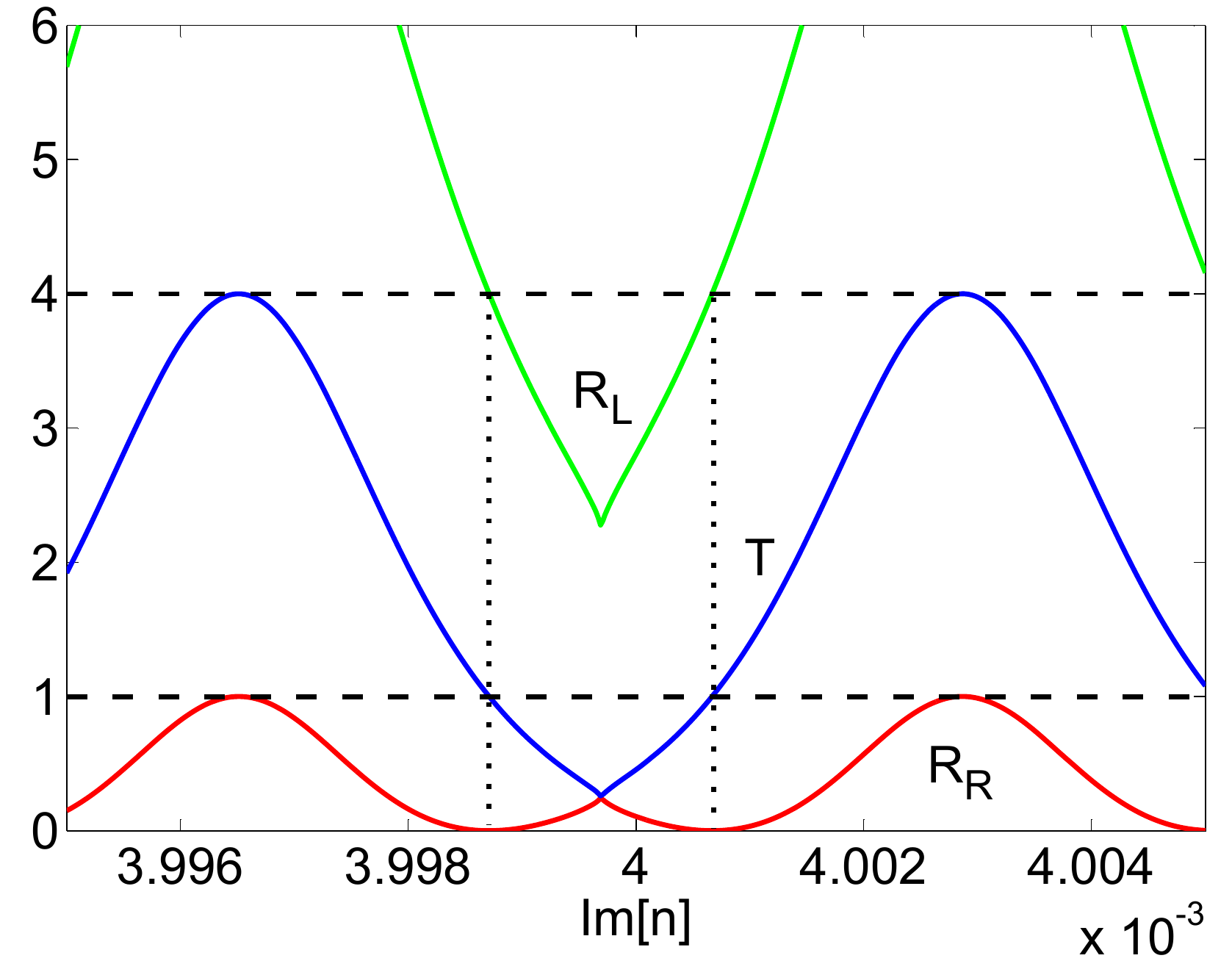}
  \caption{Reflectances and transmittance along the $\pt$ phase
    boundary for the 1D $\pt$-symmetric structure studied in
    Fig.~\ref{fig:RT}(a), in the high frequency regime ($\omega
    L\gtrsim 1725$).  The plots are given as a function of the
    gain/loss strength Im$[n]$, while the frequency $\omega$ is
    simultaneously varied to maintain the system at the phase
    boundary. Vertical dotted lines indicate points where $R_R = 0$,
    for which $R_L = 4$ and $T = 1$ as predicted by
    Eq.~(\ref{eq:borderCond}). } \label{fig:phaseboundary}
\end{figure}

Eq.~(\ref{eq:borderCond}) implies that for an ATR to coincide with the
exceptional point, the non-zero reflectance must be exactly equal to
4, which is allowed but will not occur without specific tuning.  An
example of such tuning is shown in Fig.~\ref{fig:phaseboundary}.  This
plot is obtained by tuning both the gain/loss strength (Im$[n]$) and
the frequency, to keep the system along the phase boundary, and
observing the reflectances and transmittance.  Two ATRs are found
along the phase boundary.  We note that a special set of solutions of
Eqs.~(\ref{eq:constraint}) and (\ref{eq:borderCond}) are given by
$R_R,R_L,T=(p\pm1)^2,p^2,(p\mp1)^2$, where $p$ is an arbitrary real
number. Interestingly, the maxima of $R_R,R_L,T$ in this simple
geometry are given by this set of solutions with $p=\text{Re}[n]$ in
the high frequency regime where $\text{Im}[n]\ll1$.

\section{Uniqueness of $\pt$ transition in scattering}
\label{Uniqueness of PT transition in scattering}

The generalized unitarity relations
(\ref{eq:constraint0})-(\ref{eq:constraint_phase}) hold regardless of
whether the eigenvalues and eigenvectors of the S-matrix are in the
$\pt$-symmetric or $\pt$-broken phase;  although the quantities in
the generalized unitarity relations are related to the phase of the
$\pt$ scattering system through the relation
Eq.~(\ref{eq:borderCond}). There is, however, some freedom of choice
in the definition of the 1D S-matrix, corresponding to permutation of
the outgoing channels.  The definition we used in Ref.~\cite{CPALaser}
is given in Eq.~(\ref{eq:Sdef}), which is also widely used in
mesoscopic physics \cite{Beenakker_rmp}. In this section we will
refer to the S-matrix defined in this way as $S_0$. In $S_0$ the
reflection coefficients are on the diagonal, and the outgoing channels
are related to the corresponding incident channels by time reversal,
which seems quite natural.  In particular, the time-reversal operation
$\mathcal{T}$ in this definition is represented by the complex conjugation operator.

There is, however, an alternate definition:
\begin{equation}
  \left( \begin{matrix} D\\ A\end{matrix}\right) = S_c \left( \begin{matrix} B\\ C\end{matrix}\right) \equiv \left( \begin{matrix} t &r_L \\ r_R & t\end{matrix}\right) \left( \begin{matrix} B\\ C\end{matrix}\right),
\end{equation}
which has also been used in the literature, including in one of the
earliest works on $\pt$-symmetric scattering, by Cannata {\it et
  al.}~\cite{Cannata}.  This alternative definition of the S-matrix,
which we will refer to as $S_c$, was subsequently used in the work on
unidirectional invisibility of Lin {\it et al.}  \cite{invisibility}.
Because the permutation operation does not preserve the eigenvalues,
these two different definitions of the S-matrix lead to different
criteria for the symmetric and broken symmetry phases, as well as for
the phase boundary (exceptional points).  This can lead to confusion,
as well as raising questions as to whether the S-matrix eigenvalues
and eigenvectors, and their transitions, are physically meaningful.

Note first that both definitions lead to the same values for
$t,r_R,r_L$, so they will give the same scattered state for the same
input state.  The issue is whether one or the other definition more
closely reflects the phenomena of spontaneous $\pt$ symmetry breaking,
as already known from Hamiltonian studies.  In our earlier work on the
$\pt$ transition in scattering systems \cite{CPALaser}, we showed that
the phase boundary of $S_0$ corresponds closely to the anti-crossings
of the poles of the S-matrix in the complex $\omega$-plane (see also \cite{schomerus2}).  The
locations of these poles are independent of the definition of $S$;
they reflect the internal excitation frequencies of the scatterer, as
well as the coupling of these excitations to the continuum. This
suggested that the $\pt$-transition of $S_0$ is indeed associated with
the $\pt$ transition of some underlying effective $\pt$-symmetric
Hamiltonian.  We have recently verified this point of view
analytically and numerically, in collaboration with others.  The main
part of that work will be presented elsewhere \cite{closed}; here we
just state a few relevant results and show a numerical example
corroborating this point of view.

First, it is straightforward to show that the eigenvalues of $S_c$ have the same general properties as those of $S_0$,
(even though they don't coincide).  In particular, their product is $-1$ and they are either both unimodular or
of reciprocal modulus.  However the criterion for their exceptional points differs from that of $S_0$.
Using a similar $a,b,c$ parametrization of $\Sc$ as used earlier for $S_0$, one finds that the eigenvalues are given by:
\be
\lambda_1,\lambda_2 = \frac{1}{a}\left[1 \pm \sqrt{-bc}\right] \label{eq:eigCond_Sc}.
\ee
Since both $b$ and $c$ are real, this expression shows that when $bc>0$ both eigenvalues are complex (and unimodular);
whereas when $bc<0$, both eigenvalues are real and satisfy $|\lambda_1|=|\lambda_2|^{-1}\neq1$.
Exceptional points occur when $b=0$ or $c=0$.  From Eqs.~(\ref{eq:S}) and (\ref{eq:constraint0}) one sees that $bc = (1/T-1)$ and so
$bc> 0 \to T<1$ and $bc < 0 \to T>1$, while $b=0 (r_L=0)$ or $c=0 (r_R = 0)$ is the condition for $T=1$.
{\it Thus each ATR is an exceptional point for $S_c$}, and $T>1$ corresponds to the ``broken symmetry" phase, whereas $T<1$  to the ``symmetric" phase.
This is in contrast to $S_0$ for which one has the criterion of Eq.~(\ref{eq:borderCond}) involving both $T$ and the average of $R_L$ and $R_R$.

\begin{figure}[h]
  \centering
  \includegraphics[width=\linewidth]{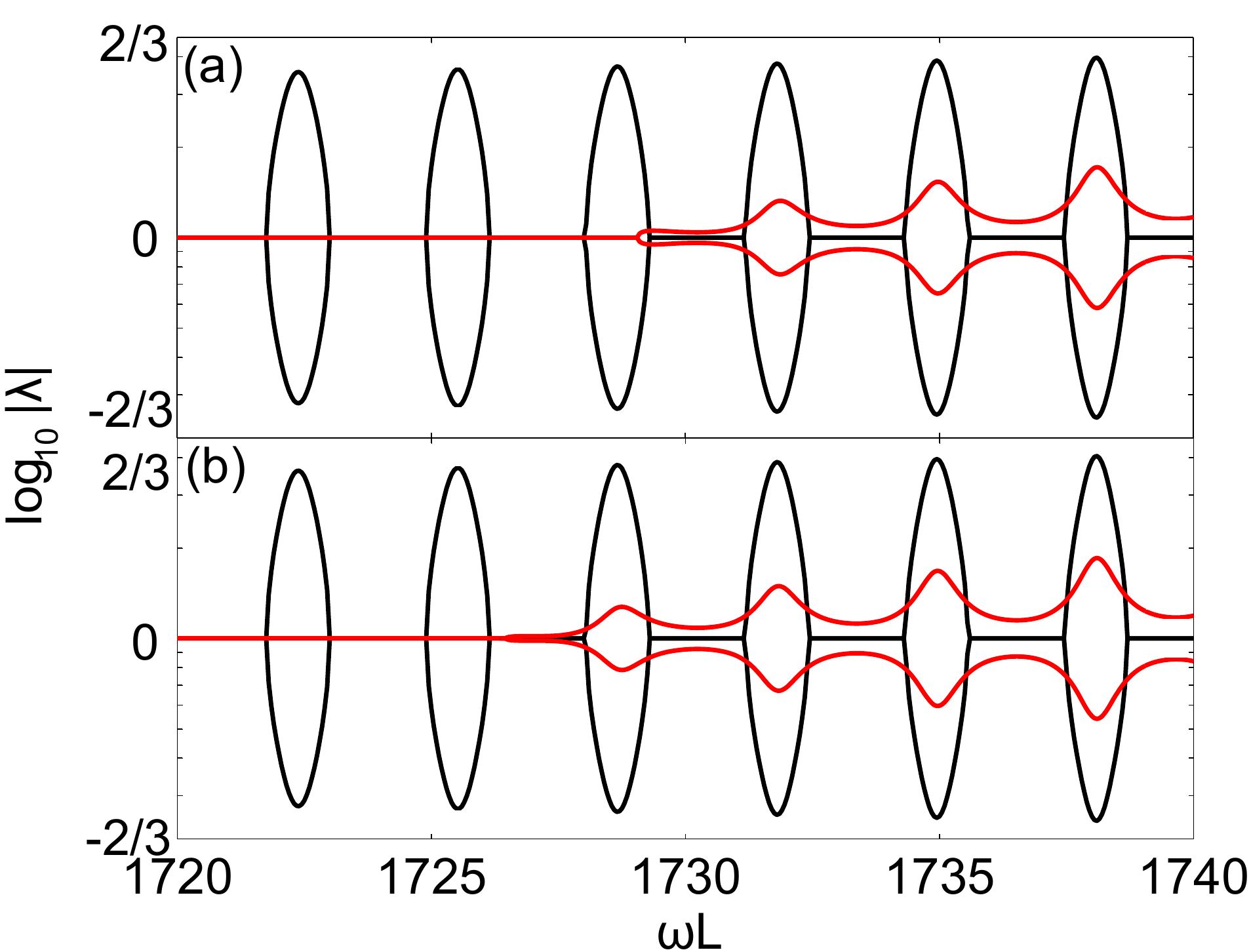}
  \caption{Logarithm of the modulus of the eigenvalues of $S_0$ (red) and $S_c$ (black) for the 1D $\pt$ heterostructure studied in Fig.~\ref{fig:RT}(a); case (a) is with Im$[n] = 3.995\times10^{-3}$ and case (b) is with Im$[n]= 4\times10^{-3}$.
   $|\lambda|=1$ indicates the $\pt$ symmetric phase, and reciprocal values for $|\lambda|$ indicate the broken symmetry phase.
   $S_c$ has multiple ``transitions" spaced by the FSR and insensitive to Im$[n]$; $S_0$ has a single transition which is
   highly sensitive to small changes in Im$[n]$. } \label{fig:phaseTransition}
\end{figure}

\begin{figure}[h]
  \centering
  \includegraphics[width=\linewidth]{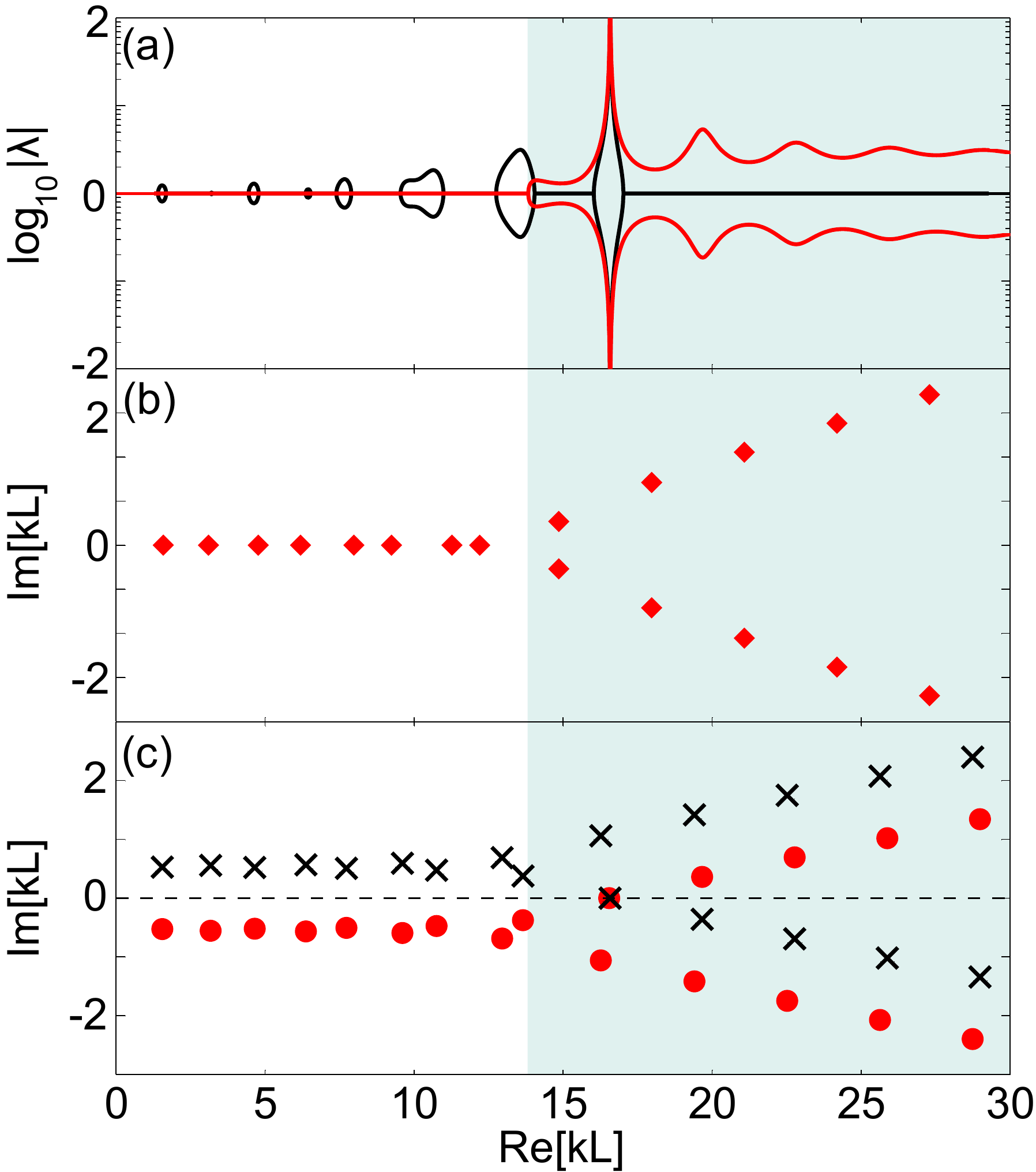}
  \caption{(a) $\pt$ phase transition of $S_0$ (red) and $S_c$ (black) for the 1D heterostructure studied in Fig.~\ref{fig:RT}(a) but with $n=2\pm0.208i$. The broken symmetry phase of $S_0$ is indicated by the shadowed area. (b) $\pt$ phase transition of the corresponding finite non-Hermitian cavity Hamiltonian with Dirichlet boundary conditions. Its broken symmetry phase coincides with that of $S_0$. (c) Poles (red circles) and Zeros (black crosses) of the S-matrix
(which are the same for $S_0$ and $S_c$); anti-crossing of the poles occurs at the phase transition of $S_0$.    At $kL \approx 17$ the system
is very near the CPA-Laser threshold point, where the cavity both emits laser radiation and perfectly absorbs the time-reverse of the lasing mode.
  } \label{fig:compClosed}
\end{figure}

These two conditions for the transition and for the two phases of the S-matrix do not coincide (see Figs.~\ref{fig:phaseTransition} and \ref{fig:compClosed}(a)) unless an ATR is tuned to occur at the phase boundary of $S_0$ as we have shown in Fig.~\ref{fig:phaseboundary}. We see that for this simple heterostructure, $S_0$ has a single transition to the broken symmetry phase (for a fixed $\text{Im}[n]$), while $S_c$ has a series of transitions
corresponding to entering and leaving the broken-symmetry phase in the
high frequency regime (Fig.~\ref{fig:phaseTransition}).  Each of these
transitions begins at one of the two ATRs and ends at the other; thus
the centers of the broken symmetry regions are spaced by the free
spectral range of the unitary cavity. These ``lozenges" of broken
symmetry phase barely change when $\text{Im}[n]$ is varied; $S_c$
repeatedly enters and leaves the symmetric phase as we tune $\omega$.
In contrast, the single transition point of $S_0$ moves substantially
to lower frequency as $\text{Im}[n]$ increases; once it enters the
broken-symmetry phase, it never re-enters the symmetric phase at any
higher frequency.  This indicates that $S_0$, not $S_c$, is measuring
the breaking of $\pt$ symmetry.

In Fig.~\ref{fig:compClosed} we show the decisive comparison. If we simply take the $\pt$ heterostructure shown in Fig.~\ref{fig:RT}a, and impose Dirichlet or Neumann boundary conditions at the boundaries to the continuum, we have a non-Hermitian discrete eigenvalue problem with $\pt$ symmetry. Its energy spectrum (expressed as complex frequencies) makes transitions between real and complex conjugate pairs (Fig.~\ref{fig:compClosed}(b)), in a manner which perfectly follows the behavior of the eigenvalues of $S_0$ and but not of $S_c$ (Fig.~\ref{fig:compClosed}(a)).
Moreover, in Fig.~\ref{fig:compClosed}(c) we show the poles and zeros of the S-matrix; their symmetric distribution around the Im$[k]$ axis is a consequence of the $\pt$ symmetry.  Before the $\pt$ transition of $S_0$ the poles have approximately the same value of Im$[k]$ as for the passive system, but just at the transition of $S_0$ there is an anti-crossing in the complex plane and half begin moving toward the real axis and the other half recede further down in the complex plane \cite{CPALaser,schomerus2}.  For Im$[k] \approx 17$ the system is very near the CPA-laser point for which
a pole and zero coincide on the real axis.  The eigenvalues of both $S_0$ and $S_c$ diverge/vanish at this point because $t,r_R$ and $r_L$ all diverge at the lasing transition. Interestingly, for this value of Im$[n]$, there are no ATRs after the lasing transition and $T<1$ for all larger $k$; the reasons for this are discussed in the Appendix.  The same correspondence between the broken symmetry phase of $S_c$ and the analogous closed system hamiltonian holds for more complex $\pt$ heterostructures, such as that of Fig.~\ref{fig:RT}(c), where $S_0$ has multiple broken phases \cite{closed}. Thus we believe that at least for the 1D case, there is a unique definition of the S-matrix, under which its $\pt$ transitions actually reflect the symmetry breaking in the underlying non-Hermitian Hamiltonian.

\section{Conclusion}
We have derived generalized unitarity relations for the S-matrix of
arbitrary 1D $\pt$-symmetric photonic heterostructures, including a
conservation relation between the transmittance and the left and right
reflectances.  This conservation relation can be easily tested in
experimental structures and used as a criterion of how precisely one
has realized the $\pt$ symmetry.  In addition, the conservation
relation leads to a simple criterion for identifying the exceptional
point(s) at which the $\pt$ symmetry is spontaneously broken or
restored. These exceptional points are shown to be closely related to
the $\pt$-symmetry breaking transition of the underlying effective
Hamiltonian of the system.

\section{Acknowledgments}

We thank Tsampikos Kottos, Stefan Rotter, Kostas Makris, Hamidreza Ramezani, Zheng Lin, Hui Cao, Demetri Christodoulides, and Carl West for helpful conversations. This work was supported by NSF grant No. ECCS 1068642.

\appendix
\section{Properties of simple gain-loss heterojunction}

The $\pt$ heterostructure shown in Fig.~\ref{fig:RT}(a), which consists of two uniform slabs of equal length and index $n,n^*$ is the simplest example one can study of the class of 1D $\pt$ symmetric photonic heterostructures, and it has been treated previously in \cite{CPALaser, mostafazadeh}.  We will refer to this structure as the ``simple heterojunction" (SH), and it is described by the transfer matrix (\ref{eq:M}) with $a = (\alpha+\alpha^*) + i(\beta+\gamma)$, $b = -i(\alpha-\alpha^*) + (\gamma-\beta)$, and $c = i(\alpha-\alpha^*) + (\gamma-\beta)$, where
\bea
\alpha &=& \frac{|\cos\Delta|^2}{2}  - \frac{n^*}{2n}|\sin\Delta|^2, \\
\beta & = & \frac{1}{2|n|^2}[n^*\sin\Delta\cos\Delta^* + c.c.], \\
\gamma &=& \frac{1}{2}[n\sin\Delta\cos\Delta^* + c.c.],
\eea
and $\Delta\equiv nkL/2$ is the complex optical path inside the left half. Since $\beta,\,\gamma$ are real, so are $b$ and $c$, and it is straightforward to check that (\ref{eq:c_constraint}) holds. This transfer matrix leads to certain simple properties. First, as mentioned in the text, the SH has no trivial ATRs as we will show in subsection \ref{sec:App1}. Second, below the $\pt$ symmetry breaking point it has many ATRs, roughly two per free spectral range of the passive resonator.  Above the symmetry breaking transition it still has ATRs until it passes the lasing transition after which they disappear in the limit $\text{Im}[nkL] \to \infty$.  We will discuss this behavior and contrast it with more complex heterostructures in subsection \ref{sec:App2}.

\subsection{Absence of trivial ATRS}
\label{sec:App1}
The SH can be treated as having an air gap of vanishing width in between the gain and loss regions. Hence the absence of trivial ATRs is a consequence of the absence of reflectionless transmission resonances of such uniform amplifying or attenuating slabs in air. Below we first discuss in general the transmission resonances of a uniform slab of refractive index $n$ and length $L/2$ embedded in two semi-infinite media of index $n_l$ and $n_r$.

\begin{figure}[h]
  \centering
  \includegraphics[width=0.8\linewidth]{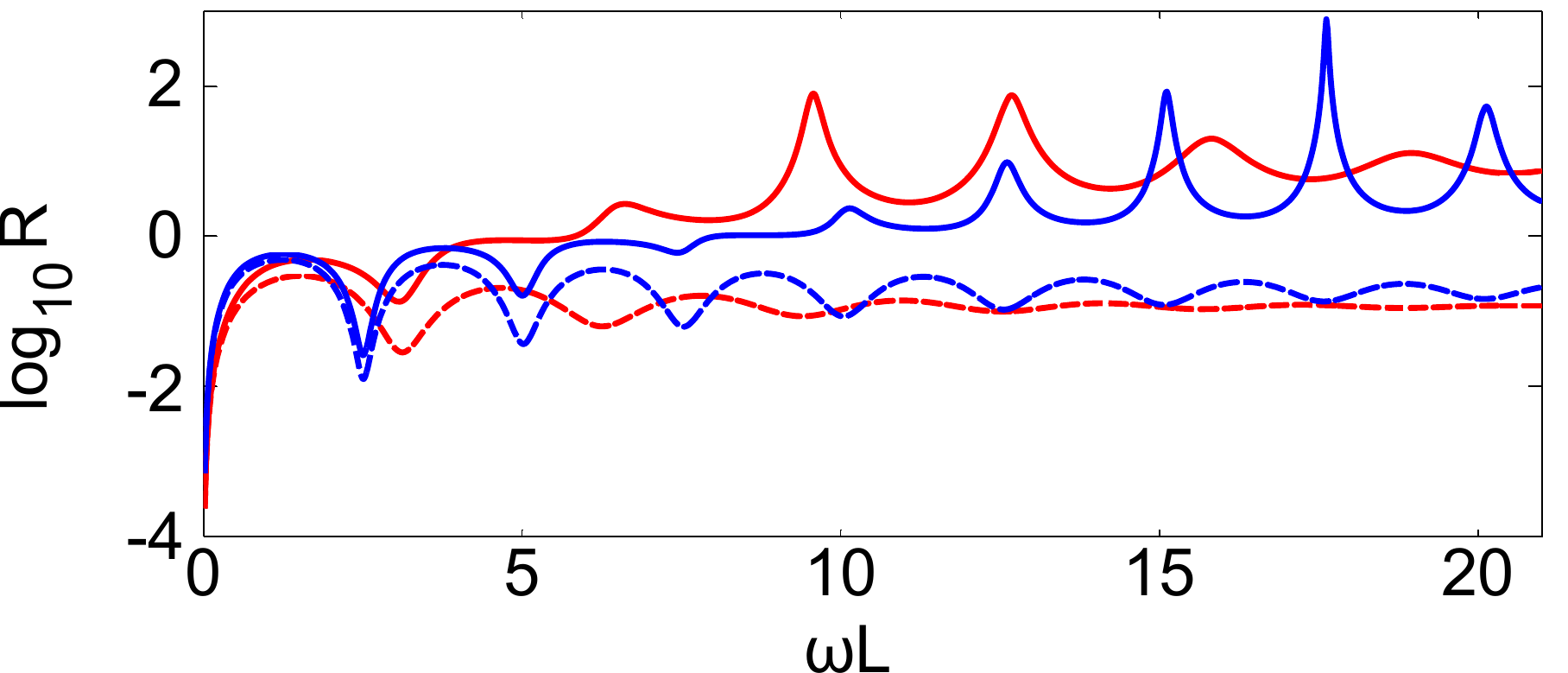}
  \caption{Reflectance of the gain (solid) and loss (dashed) region in the heterostructure studied in Fig.~\ref{fig:RT}(a). Two values of the refractive index are considered: $n=2\pm0.2i$ (red), $n=2.5\pm0.1i$ (blue). Solid and dashed lines indicate $R_L,\,R_R$ of the gain region and the loss region, respectively. } \label{fig:RT_single}
\end{figure}

For this simple setup the transfer matrix defined in Sec.~II ($\left( \begin{smallmatrix} A\\ B\end{smallmatrix}\right) = M \left( \begin{smallmatrix} C\\ D\end{smallmatrix}\right)$; see Fig.~\ref{fig:schematic}) takes the form
\be
M = \frac{1}{2}\left(\begin{array}{c c}
        1 & \frac{1}{n_lk} \\
        1 & -\frac{1}{n_lk}
        \end{array}\right)
        \left(\begin{array}{c c}
        \cos\Delta & i\frac{\sin\Delta}{nk} \\
        ink\sin\Delta & \cos\Delta
        \end{array}\right)
        \left(\begin{array}{c c}
        1 & 1 \\
        n_rk & -n_rk
        \end{array}\right),\label{eq:M1}
\ee
where $n_l,\,n_r,\,n$ can be complex. A transmission resonance of an incident beam from the left side requires $C=0$ and
\be
A = \left(1-\frac{n_r}{n_l}\right)\cos\Delta + i\left(\frac{n}{n_l}-\frac{n_r}{n}\right)\sin\Delta = 0. \label{eq:transRes}
\ee
For the gain and loss regions in SH, when treated as being separated by an infinitely thin air gap, $n_l=n_r=1$ while $\text{Im}[n]\equiv \tau \neq0$. We immediately see that Eq.~(\ref{eq:transRes}) cannot be satisfied because $\sin\Delta\neq0$ due to the finite imaginary part of $n$. This holds independent of $n$ ($\tau\neq0$) and $k(\neq0)$ (see Fig.~\ref{fig:RT_single}). This finding is confirmed by calculating the reflectance directly (see Fig.~\ref{fig:RT_single}(a)). The same analysis can be extended to the slightly more complicated case shown in Fig.~\ref{fig:RT}(c), where all the ATRs are also found to be ``non-trivial'' as confirmed again by calculating the reflectances of the sub-units directly. We note, however, that trivial ATRs do exist in some $\pt$ structures. An example is the concatenation of even numbers of SHs.

\subsection{Asymptotic properties of ATRs in the simple heterojunction}
\label{sec:App2}
\begin{figure}
  \centering
  \includegraphics[width=0.8\linewidth]{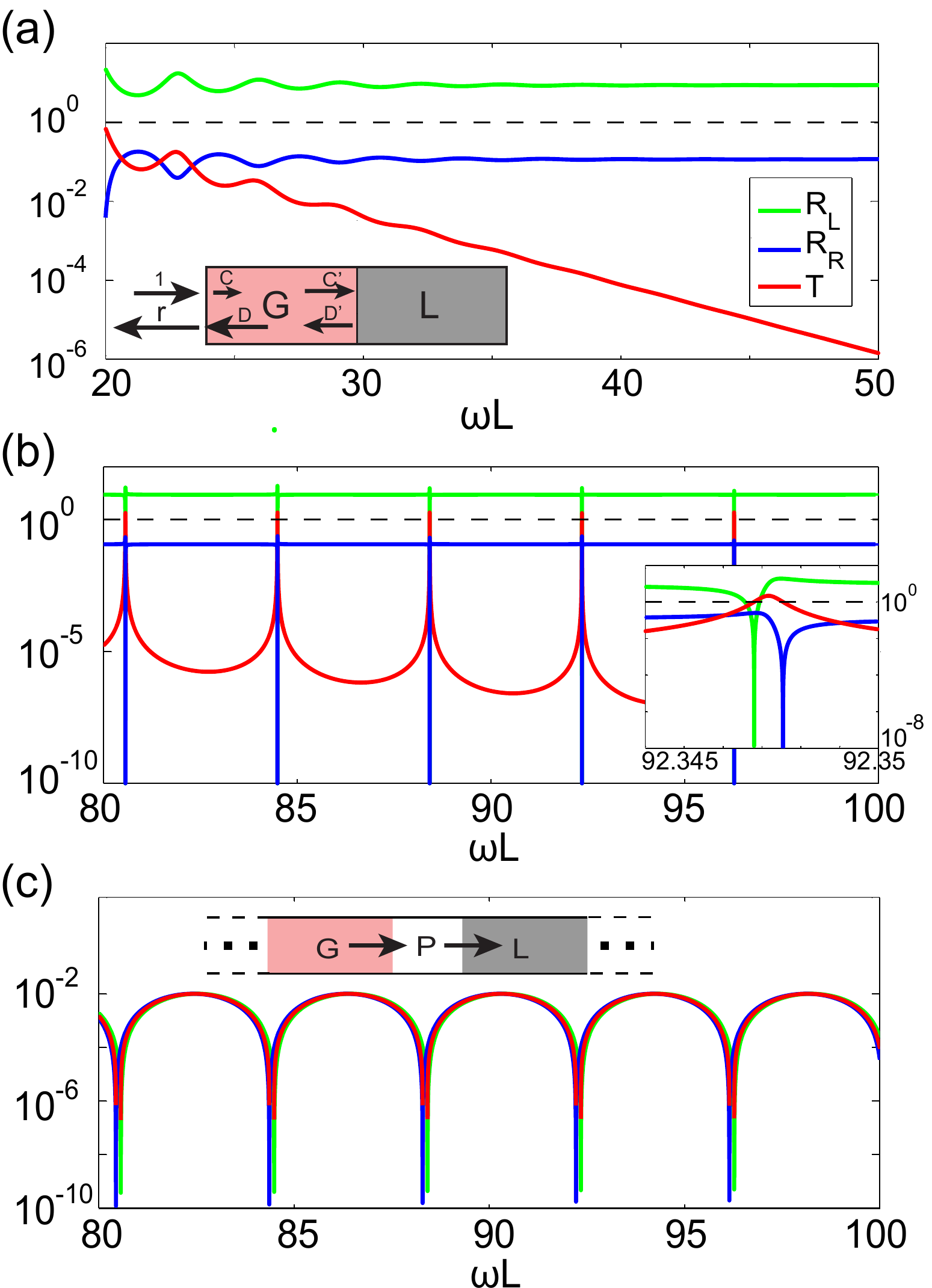}
  \caption{(a) Reflectances and Transmittance same as in Fig.~\ref{fig:RT}(a) but at higher frequencies. Inset: Analysis of the reflection coefficient from the ``gain'' side. Letters with arrow represent the complex amplitude of the traveling waves at the nearest interface. (b) Reflectances and Transmittance same as in Fig.~\ref{fig:RT}(c) but at higher frequencies. Inset: zoomed in on the two ATRs near $\omega{L}=92.35$. (c) Reflectances and transmittance of the central passive region in Fig.~\ref{fig:RT}(c) placed between two semi-infinite regions of gain and loss with $n=2\pm0.2i.$ The transmission resonances are given by Eq.~(\ref{eq:GNL_transRes}) in which $\text{Im}[n_l]$ takes opposite signs depending on the propagation direction. Red curve represents $T-1$ and its broken parts indicate $T<1$.
  } \label{fig:r_gain}
\end{figure}

In this main text we noted that ATRs for the SH disappear soon after the lasing threshold is passed in the broken symmetry phase of $S_0$ (see Fig.~\ref{fig:r_gain}(a)). To understand this observation, we study the behaviors of $R_R$ and $R_T$ in the large $\tau kL$ limit. The reflection coefficient from the ``loss'' side of the $\pt$ heterostructure approaches the value given by the Fresnel formula, $(1-n)/(1+n)$, due to the suppression of interference effects by strong absorption. Surprisingly, the asymptotic value of the reflection coefficient from the ``gain'' side, $(1+n^*)/(1-n^*)$, turns out to be the inverse of the Fresnel formula. This can be explained from the analysis shown by the inset of Fig.~\ref{fig:r_gain}(a). Due to the strong loss inside the ``loss'' side of the cavity, $D'$ is given by the Fresnel relation $C'(2n^*)/(n^*-n)\equiv C'r'$, implying $|D|=|\exp[in^*kL/2]D'|=|r'\exp[in^*kL]C|$ is much larger than $|C|$ in the large $\tau kL$ limit. Therefore, the scattering at the air-gain interface is approximately the time reversed process as if the ``gain'' side (which is the ``loss'' side in the time-reversed picture) were semi-infinite, i.e. with incident amplitude $r^*$ and reflected amplitude $1$ in the air and transmitted amplitude $D^*$, satisfying the Fresnel relation. The reflection coefficient in the original problem is then $(1+n^*)/(1-n^*)$. Thus $R_LR_R\rightarrow1$ in this limit, which implies $T\rightarrow0$ from Eq.~(\ref{eq:constraint}) and ATRs do not exist.

In more complicated $\pt$ structures ATRs can take place in this limit. For example, the reflection coefficient connecting $C'$ and $D'$ approaches zero at the transmission resonances of the passive region in Fig.~\ref{fig:RT}(c). The analysis above then breaks down and sharp changes of the transmittance and reflectances take place at these frequencies as shown in Fig.~\ref{fig:r_gain}(b). These transmission resonances through the passive region is a special set of solutions of Eq.~(\ref{eq:transRes}). They require $n_l=n_r^*$ and $\text{Im}[n]=0$, and the transmission resonances occur at
\be
\Delta = \arctan\left[\frac{2\text{Im}[n_l]n}{\text{Re}[n_l]^2-\text{Im}[n_l]^2-n^2}\right].\label{eq:GNL_transRes}
\ee
In the frequency range shown in Fig.~\ref{fig:RT}(c) where $\text{Im}[n_l]k$ is small, these transmission resonances do not lead to ATRs of the whole heterostructure due to the multiple interferences taking place inside the gain and loss sub-units. In the large $\text{Im}[n_l]k$ limit shown in Fig.~\ref{fig:r_gain}(b), however, these multiple interferences are suppressed due to strong absorption/amplification, and ATRs arise from the resonances given by (\ref{eq:GNL_transRes}). Note that these ATRs are still ``non-trivial'' as the frequencies given by (\ref{eq:GNL_transRes}) are not the transmission resonances of the gain or loss sub-unit in the absence of the other.

For the purpose of completeness, we mention a few more cases where transmission resonances of a single uniform slab (i.e. the solutions of Eq.~(\ref{eq:transRes})) exist. When $n_l,\,n_r\,,n$ can be treated as real (with negligible absorption and no gain), two types of solutions of (\ref{eq:transRes}) can be found. The first one is well-known, $n_l=n_r$, which requires $\sin\Delta=0$; the second one is less well-known, $n=\sqrt{n_ln_r}$, which requires $\cos\Delta=0$. It is easy to convince oneself that no other types of solution exist for real indices. As one slowly increase the gain or loss strength in the scattering layer, approximate transmission resonances can still be found, but their reflectances gradually increase and eventually become detectable. In Ref.~\cite{surfaceMode} a different case was studied where $n_r=1, \text{Im}[n]=0,\,\text{Im}[n_l]\neq0$. By noticing that $\cos\Delta=0$ cannot satisfy the above equation and $\tan\Delta$ is real, Eq.~(\ref{eq:transRes}) can be reduced to
\bea
\text{Im}[n_l]^2 &=& (\text{Re}[n_l]-1)(n^2-\text{Re}[n_l]), \label{eq:transRes_surfaceMode}\\
\tan\Delta &=& -n\left(\frac{\text{Re}[n_l]-1}{n^2-\text{Re}[n_l]}\right)^\frac{1}{2}
\eea
It describes the transmission resonance from a loss/gain media to air through a passive slab, which gives rise to the novel ``surface'' lasing modes introduced in Ref.~\cite{surfaceMode}.


\begin{thebibliography}{99}
\bibitem{bibnote0} When $M_{11},M_{22}=0$ the two incident amplitudes $B,C$ and the two outgoing amplitudes $A,D$ become decoupled, which gives rise to the CPA-laser points and the reflection and transmission coefficients become infinite. However, the results derived in the main text are still valid at these points with appropriate choices of limiting procedures.
\bibitem{bibnote:accidental_invisibility} There may seem to be a number of frequencies in Fig.~\ref{fig:RT}(b) at which $T=1$ and $\phi_t  \approx 0$, but it is a coincidence as the transmission phase oscillates around $nkL$ and we have chosen $\text{Re}[n]$ to be an integer.
\end{thebibliography}
\end{document}